\documentclass{ws-p9-75x6-50}

\begin{document}

\title{Brueckner Theory of Nuclear Matter with Nonnucleonic Degrees
of Freedom and Relativity
\thanks{\rm Invited talk presented at the 
{\it Tenth International Conference on Recent Progress in Many-Body Theories},
September 10--15, 1999, Seattle, Washington, USA;
to be published in {\it Advances in Quantum Many-Body Theory},
Vol.~3 (World Scientific, Singapore).}
\thanks{\rm Dedicated 
to Keith Brueckner on the occasion of his 75th birthday.}}

\author{R. Machleidt}

\address{Department of Physics, University of Idaho, Moscow, ID 83844, USA
 \\E-mail: machleid@uidaho.edu}

\maketitle

\abstracts{
For the past 40 years, Brueckner theory has proven to be
a most powerful tool to investigate systematically models for
nuclear matter. I will give an overview of the work 
done on nuclear matter theory, starting with the simplest
model and proceeding step by step to more sophisticated
models by extending the degrees of freedom and including relativity.
The final results of a comprehensive hadronic theory of
nuclear matter are compared to the predictions by currently
fashionable two-nucleon force models. It turns out that a
two-nucleon force can, indeed, reproduce those results
{\it if the potential is nonlocal}, since nonlocality is an inherent
quality of the more fundamental fieldtheoretic
approach. This nonlocality
is crucial for creating sufficient nuclear binding.
}

\section{Introduction}

By definition, {\it nuclear matter\/} refers to an infinite uniform system
of nucleons interacting via the strong force without electromagnetic
interactions. This hypothetical system is supposed
to approximate conditions in the interior of a heavy nucleus.
We shall assume equal neutron and proton densities;
that is, we will consider symmetric nuclear matter.
This many-body system is charactarized by its energy per nucleon as a function
of the particle density. Based upon various semi-empirical sources,
nuclear matter is determined to saturate at a density
$\rho_0 = 0.17 \pm 0.02$ fm$^{-3}$ (equivalent to a Fermi
momentum $k_F = 1.35 \pm 0.05$ fm$^{-1}$)
and an energy per nucleon ${\cal E}/A = -16\pm 1$ MeV.

Historically, the first nuclear matter calculations were performed 
by Heisenberg's student Hans Euler, in 1937.\cite{Eul37}
This was just two years after Weiz\"acker\cite{Wei35} had suggested the
semiempirical mass formula. Euler applied an attractive potential 
of Gaussian shape in second-order perturbation theory.

Modern studies began in the early 1950's after a repulsive core
in the nuclear potential had been conjectured.\cite{Jas51}
It was obvious that conventional perturbation theory was inadequate 
to handle such singular potentials. Therefore, special methods had to be
developed. This program was initiated by {\it Keith Brueckner\/} and 
co-workers\cite{BLM54,Bru54,BL55} who applied---for the 
ground-state problem of nuclei---methods similar to those
developed by Watson\cite{Wat53} for multiple scattering. Later,
a formal basis for this new approach was provided by Goldstone\cite{Gol57}
who, using perturbation theoretical methods, established the so-called linked
cluster expansion. The success of Brueckner theory in practical
calculations stems from the fact that certain classes of linked
diagrams can be summed in closed form up to infinite orders defining
the so-called reaction matrix $G$. All quantities are then formulated 
in terms of this $G$ which---in contrast to the bare nuclear 
potential---is smooth and well behaved.

The first numerical calculations applying Brueckner theory were performed
in 1958 by Brueckner and Gammel\cite{BG58} using the Gammel-Thaler
potential.\cite{GT57}
In the 1960s, Hans Bethe and his students 
elaborated thoroughly on
Brueckner theory.\cite{Day67,RB67,Bra67,Bet71} 
Around 1970, elegant numerical methods\cite{Sie70,HT70}
for the solution of the Brueckner equation were established
and sytematic calculations performed.\cite{Coe70,Spr72}

In the mid 1970s, the nuclear many-body community was shaken by
an apparent discrepancy between results from Brueckner theory
and the variational approch (the famous/infamous
`nuclear matter crisis'\cite{Cla75}). This indicated that both many-body
theories had to be reaxamined and more consistent calculations
had to be performed. 
For Brueckner theory, the necessary work was done mainly
by Ben Day.\cite{Day78,Day81} 
The variational approach was pursued by the Urbana
group,\cite{PW79} the Rome-Pisa collaboration,\cite{FR75} and
many others.\cite{MB1} 
As a result of this enormous work, quantitatively 
very close
predictions were finally obtained from the two different many-body
approaches using realistic $NN$ potentials.\cite{LP81,DW85,MB4}
Based upon these comprehensive investigations,
it is now commonly accepted that
both many-body approaches are reliable for densities typical
for conventional nuclear physics.

\begin{figure}[t]
\hspace*{3cm}
\epsfig{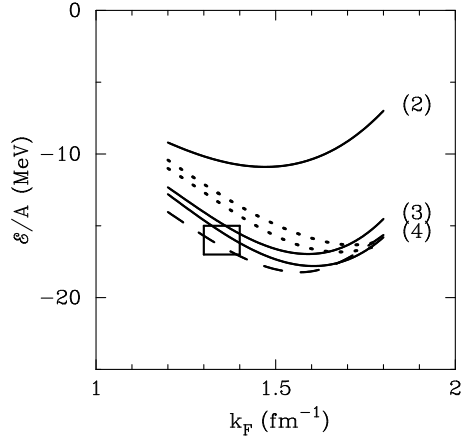}
\caption{Energy per nucleon in nuclear matter, 
${\cal E}/A$, 
{\it versus\/} the Fermi momentum, $k_F$, applying the Argonne
$V_{14}$ potential. Solid curves represent Brueckner results,
using the gap choice for the single particle potential,
with the number of hole lines indicated in parentheses.
The dashed curve is the Brueckner two hole-line result
using the continuous choice.
The two dotted lines are obtained in variational calculations.
The box describes the area in which nuclear saturation
is expected to occur empirically.
(Dashed curve from Ref.~$^{30}$, all other curves from
Ref.~$^{26}$.)}
\end{figure}

Brueckner theory does not converge in powers of $G$, but in terms
of the hole-line expansion.\cite{Bra67}
This is demonstrated in Fig.~1 for the case of the Argonne $V_{14}$
potential.\cite{WSA84} 
The solid lines in this figure 
are obtained from Brueckner theory
using the conventional choice (`gap choice')\cite{foot1} for the single
particle potential;
the number of hole lines taken into account
are indicated in parentheses. 
The Brueckner results shown in Fig.~1 are from 
Day;\cite{DW85} they were recently confirmed
by Baldo and coworkers.\cite{Son98}
The dotted lines in Fig.~1 are the predictions by
two variational calculations performed by Wiringa.\cite{DW85}
Within the uncertainties there appears to be
agreement between Brueckner theory and the variational approach;
particularly, around the minimum of the curves.
Finally, the dashed curve in Fig.~1 is the lowest order Brueckner
(i.e., two hole-line) result when the continuous choice\cite{foot1} is
used for the single particle potential.\cite{Son98}
It is clearly seen that this latter result agrees well
with the four hole-line Brueckner prediction using
the gap choice.
This may suggest that the continous choice
is a more efficient way to perform Brueckner 
calculations.\cite{Son98,JLM76,GL79,Mah79}

\begin{figure}[t]
\hspace*{3cm}
\epsfig{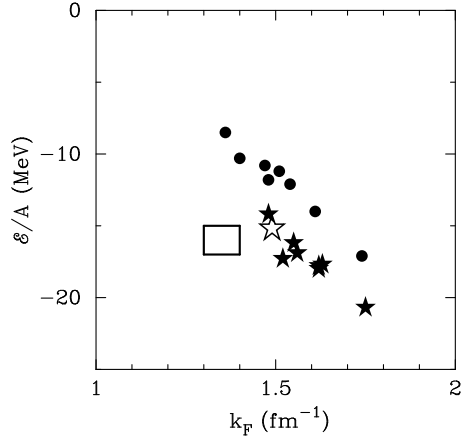}
\caption{Nuclear matter saturation as predicted by a variety of
$NN$ potentials. Solid dots are the saturation points 
(i.~e., minima of ${\cal E}/A$ {\it versus\/} $k_F$ curves)
obtained in the Brueckner two hole-line approximation
applying the gap choice for the single particle potential.
Solid stars denote corresponding predictions that either
include four hole-lines using the gap choice or two hole-lines
using the continuous choice.
The open star is the result by Brueckner and Gammel of
1958.$^{9}$}
\end{figure}

Systematic Brueckner calculations conducted with a variety of $NN$
potentials show the so-called Coester Band\cite{Coe70} structure (Fig.~2):
no prediction is able to reproduce the empirical saturation density
{\it and\/} energy simultaneously. If the density is predicted right
the energy is too high, and if the energy is right
the density is too high.
The predictions discussed so far
were obtained in the framework of the simplest model for the atomic
nucleus: 
point nucleons obeying the nonrelativistic Schr\"odinger equation
interact through a static two-body potential that fits low-energy $NN$
scattering data and the properties of the deuteron.
This conventional model for the nucleus is obviously insufficient to
quantitatively describe the nuclear ground state.
It is, therefore, necessary to go beyond the conventional scheme.

To get some structure into the upcoming considerations,
 we list in Table 1 four items relevant to the nuclear
 many-body problem. The assumptions made in the
conventional approach are given in  column two.
Column three states some obvious ideas leading beyond
the conventional model.
Notice that the distinction between the various items
and assumption is artificial and that there is a great deal
of overlap among them.
Step by step, we will now include most of the possible extensions
in nuclear matter calculations.
One great advantage of Brueckner theory is that
it allows to take into account additional degrees of freedom
in a fairly easy and straightforward way.

\section{Meson Degrees of Freedom}

Ever since the meson hypothesis was formulated, it was
(at least in principal) clear
that the full nuclear many-body problem
should include nucleons {\it and\/} mesons.
Nevertheless, traditionally only nucleons  have been considered,
these interacting via a static two-body potential.
Even in cases where the two-nucleon force was derived from
meson theory, the mesons were usually ``forgotten'' as soon as
the nuclear force was constructed. ``Meson theory''
was merely used to provide a suitable ansatz for the two-body
potential with a convenient parametrization in terms of
masses and coupling parameters. Thus, the dynamical
presence of the mesons was ignored.
Obviously, from a more fundamental point of view, this is not
satisfactory.

Therefore, we will now include meson degrees of freedom in Brueckner
theory following the approach suggested by Sch\"{u}tte.\cite{Sch74}
This approach is fieldtheoretic in nature, treating
ba\-ry\-ons and mesons {\it a priori\/}
on an equal footing. However, a principal problem
of every field-theoretic many-body
theory is how to take into account
the effects of the many-body environment on
 the particles  and their
interactions
(e.\ g.\ the single-particle energies in the medium,
propagation in the medium, etc.).
This is difficult to do  in a covariant way.
Therefore, Sch\"{u}tte suggested to use time-ordered (``old-fashioned'')
perturbation theory,\cite{Sch61}
which is similar to the usual perturbation
theory of ordinary quantum mechanics. Thus, methods familiar from
non-relativistic many-body theories can be used.

\begin{table}[t]
\centering
\small
\caption{{\bf Basic assumptions
underlying the nuclear many-body problem}}
\begin{tabular}{lll}
\hline\hline
\\ {\it Item}     &   {\it Simplest assumption}  & {\it Possible extension(s)}\\
\\ \hline
                     &                          & Mesons,\\
{\bf Degrees of freedom} & Nucleons only            & Isobars,\\
                     &                          & Quarks and gluons\\ \\
{\bf Hadron structure} & Point structure          & (Quark) sub-structure\\ \\
{\bf Interaction(s)} & Static, instantaneous    & Non-static interactions, \\
                     & two-body potential       & Many-body forces \\ \\
{\bf Dynamical equation}   & Non-relativistic       & Relativistic\\
                     & Schr\"{o}dinger equation & Dirac equation\\
\hline\hline
\end{tabular}
\end{table}

Starting from a field-theoretic Hamiltonian for mesons and nucleons
\begin{equation}
h=t+W \, ,
\end{equation}
where $t$ denotes the unperturbed
Hamiltonian (i.~e., the operator for the free relativistic energies 
of nucleons, $E_m=\sqrt{M^2+{\bf q}_m^2}$, and mesons, $\omega_\alpha=
\sqrt{m_\alpha^2+{\bf k}_\alpha^2}$) and
$W$ the meson-nucleon interactions, 
and applying time-ordered perturbation theory, the lowest order contribution
to free-space two-nucleon scattering is of second order, namely,
\begin{equation}
V(z)=W\frac{1}{z-t+i\epsilon}W \, ,
\end{equation}
with $z$ the relativistic free energy of the
two interacting nucleons (i.~e., $z=E_m+E_n$ for nucleon $m$ and $n$).
Equation (2) can be understood as an one-boson-exchange (OBE) 
``quasi-potential'' (which is energy dependent).
The Lippmann-Schwinger equation for free-space two-nucleon
scattering is
\begin{equation}
{T}(z)=V(z)+V(z)\frac{1}{z-t+i\epsilon}{T}(z) \, .
\end{equation}

Turning to the nuclear many-body problem, 
we introduce the single-particle potential
$U$ and rewrite the Hamiltonian
Eq.~(1) by
\begin{equation}
h=h_0+h_1
\end{equation}
with the unperturbed Hamiltonian
\begin{equation}
h_0=t+U
\end{equation}
and the perturbation
\begin{equation}
h_1=W-U \, .
\end{equation}
As a consequence of this, the one-boson exchange, Eq.~(2), 
is modified in the nuclear medium where it reads
\begin{equation}
\bar{V}(\bar{z})=W\frac{1}{\bar{z}-h_{0}}W \, .
\end{equation}
Similar to conventional Brueckner theory,
the equation for the Brueckner $G$ matrix is
\begin{equation}
\bar{G}(\bar{z})
=\bar{V}(\bar{z})+\bar{V}(\bar{z})\frac{Q}{\bar{z}-h_{0}}\bar{G}(\bar{z})
\, .
\end{equation}
This equation differs from the (free-space) Lippman-Schwinger
equation (3) by the Pauli projector $Q$ which projects onto
unoccupied two-nucleon states (giving rise to the Pauli effect)
and by the energy denominator $[\bar{z}-h_0]$ which replaces
$[z-t+i\epsilon]$ of Eq.~(3) (causing the dispersion effect).
Both these effects are the crucial saturation mechanisms
of conventional Brueckner theory.
Note that in all results presented in this paper the conventional
Brueckner effects are always included.
In lowest order, the energy per nucleon in nuclear matter is given by
\begin{equation}
\frac{{\cal E}}{A}=\frac{1}{A}
\sum_{m\leq k_{F}} E_{m}
+ \frac{1}{2A} \sum_{m,n\leq k_{F}}
\langle mn|\bar{G}(\bar{z})|mn-nm\rangle -M
\end{equation}
with $M$ the mass of the free nucleon.
The single particle potential used in these calculations is defined by
\begin{equation}
U(m)=Re \sum_{n\leq k_{F}}
 \langle mn|\bar{G}(\bar{z})|mn-nm\rangle 
\end{equation}
which is applied for nucleons below and
above the Fermi surface (continuous choice). 
In the nuclear medium,
\begin{equation}
\bar{z}=\epsilon_{m}+\epsilon_{n} \, ,
\end{equation}
with
\begin{equation}
\epsilon_{m}=E_m+U_m \, ,
\end{equation}
in accordance with the unperturbed Hamiltonian $h_0$.

Formally, Eqs.~(8)--(10) are essentially identical to ordinary
Brueckner theory. The difference is in the quasipotential $\bar{V}$
which depends on the nuclear medium (and its density).
The impact of this medium effect is demonstrated in Fig.~3
where the dash-triple-dot curve is calculated with this effect
while the dotted curve is without it.
It is seen that the medium effect on meson propagation 
is such that the binding energy per nucleon is
slightly reduced (in the order of 2 MeV at nuclear matter density).
This effect is similar to the dispersion effect
of ordinary Brueckner theory, but, much smaller. 
The density-dependence of this mesonic effect is such that
the saturation point moves along the Coester band and
not off it.

\begin{figure}[t]
\hspace*{3cm}
\epsfig{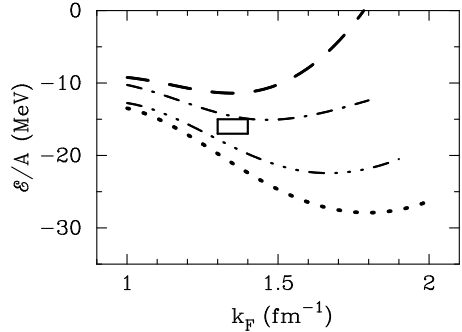}
\caption{Results from nuclear matter Brueckner calculations
that include medium effects on the two-nucleon interaction.
The dotted line is obtained without such medium effects;
the dash-triple-dot curve includes the medium effect
on meson propagation; in addition to this, the dash-dot
curve contains the dispersion effect from diagrams
with intermediate $\Delta$ states. Finally, the dashed
curve includes all medium effects (i.~e., also the
Pauli effects from $N\Delta$ intermediate states).
All calculations are conducted in the two hole-line approximation
using the continuous choice.}
\end{figure}

\section{Iso\-bar Degrees of Freedom}

\begin{figure}[t]
\vspace*{-4.0cm}
\hspace*{-3.5cm}
\epsfig{file=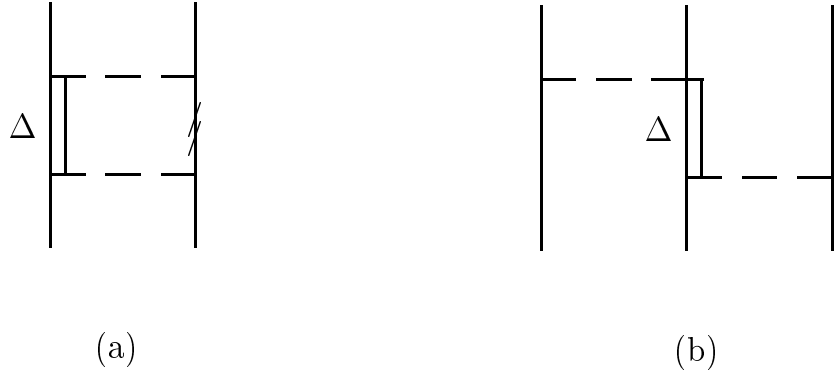,width=20.0cm}
\vspace*{-16.5cm}
\caption{Two- and three-body forces created by the $\Delta$ isobar.
Solid lines represent nucleons, double lines $\Delta$ isobars,
and dashed lines $\pi$ and $\rho$ exchange.
{\bf Part (a)} is a contribution to the two-nucleon force.
The double slash on the intermediate nucleon line is to indicate
the medium modifications (Pauli blocking and dispersion effects)
that occur when this diagram is inserted into a nuclear many-body
environment.
{\bf Part (b)} is a three-nucleon force (3NF).}
\end{figure}

The excited states of the nucleon play an important
role in a genuine and realistic meson-exchange model for the
nuclear force.
The lowest-lying pion-nucleon resonance, the  $\Delta$(1232)
isobar, is essential for $NN$ scattering at low
and intermediate energies.
It provides a large part of the intermediate range attraction
and it generates most of the inelasticity above pion-production threshold.
Again, being aware of how crucial this degree of freedom is for the
two-body interaction,
we should not ``freeze it out'' in the many-body problem.

The lowest order in which isobars can contribute to the
$NN$ interaction is the fourth order in the interaction
Hamiltonian $W$ (which corresponds to a two-meson exchange).
The general structure of the fourth order perturbation is
for the case of free scattering
\begin{equation}
V^{(4)}(z)=W \frac{1}{z-t+i\epsilon}
 W\frac{1}{z-t+i\epsilon}
 W\frac{1}{z-t+i\epsilon} W
\end{equation}
with $W$ the meson baryon interaction. The irreducible part of these
diagrams contributes to the ``kernel'' $V(z)$ of the scattering
equation. 
One of the diagrams generated by Eq.~(13) is shown in Fig.~4(a).
When inserted into the many-body problem
the contribution $V^{(4)}(z)$ is altered in a characteristic
way; namely it is replaced by
\begin{equation}
\bar{V}^{(4)}(\bar{z})=W \frac{Q}{\bar{z}-h_{0}}
 W \frac{Q}{\bar{z}-h_{0}}
 W \frac{Q}{\bar{z}-h_{0}}W
\end{equation}
where the Pauli operator $Q$ projects nucleons onto unoccupied nucleon
states.

One can distiguish between  two ways in which the medium
exercises influence:
\begin{itemize}
\item The Pauli projector $Q$ cuts out the lower part of the
 nucleon spectrum in intermediate states; this leads to the
 so-called {\it Pauli effect}.
\item The propagator $[z-t+i\epsilon]^{-1}$ is replaced
 by $[\bar{z}-h_{0}]^{-1}$; the effect caused by this replacement
 has become known as {\it dispersion effect}.
\end{itemize}
Both effects reduce the absolute size of the diagram.
 Thus, for an attractive
diagram there is a net repulsive medium effect.
Notice that these two effects are similar to the conventional
Brueckner effects discussed below Eq.~(8).

Figure~3 includes the results employing the field-theoretic model
just sketched. 
The inclusion of the dispersive effects under consideration moves us
from the dash-triple-dot curve to the dash-dot curve,
and the Pauli effects bring us further up to the dashed curve.
This group of medium effects are substantially
larger than those from the OBE terms discussed in the previous section.

The results presented in Fig.~3 are based upon the Bonn Full Model for the
$NN$ interaction.\cite{MHE87} This model includes also
the (repulsive) $\pi \rho$ diagrams 
(for which the medium effect causes a net attraction). However, since
the sum of 2$\pi$ and $\pi \rho$ diagrams is attractive, the
 2$\pi$-exchange
 being dominant at intermediate range, the net medium effect
is repulsive. It is clearly seen that dispersion and Pauli effects
are about equally important, the latter  typically increasing
more strongly with  density. We note that these calculations also
include all non-iterative diagrams (stretched and 
crossed box diagrams) involving $\Delta$ isobars which
contribute about as much to the medium effects
as the box (iterative) diagrams. 

The density-dependence of the effects due to $\Delta$ degrees of freedom
is only slightly stronger than that of the conventional saturation
mechanisms, bringing the saturation point not markedly off the
Coester band.

\section{Many-Body Forces}

When discussing  many-body forces, caution is in place.
Notice that
the diagram Fig.~4(b) is a genuine three-body force in a model
space which consists of nucleon states only. In an extended
Hilbert space which includes $\Delta$ isobar states, Fig.~4(b)
represents just a three-particle correlation.
Thus, most many-body forces are artificially
created by freezing out degrees of freedom. In this respect
they are merely artefacts of the particular theoretical
framework applied.
Because of this model dependence of the terminology, it is useful
to introduce an operating definition for  $n$-{\it nucleon forces\/}
which we will take to be the following:
 forces that depend in an irreducible way on
the coordinates or momenta of $n$ {\it nucleons\/} when {\it only
nucleon degrees of freedom\/} are taken into account.

The fact that most many-body forces are an artifact of theory
implies the following rule for how to deal
with the many-body-force issue in a proper way:
when you introduce a new degree of freedom, 
take it into account in the {\it two- and many-body problem}, 
consistently. 
Field-theoretic models for the $2\pi$-exchange contribution to
the $NN$ interaction require the $\Delta(1232)$ isobar, which
also creates $n$-nucleon forces in the many-body system.
Consequently, when introducing the isobar degree of freedom, it should
be included in the two- and many-nucleon forces simultaneously.
In Fig.~4(a) we showed a contribution to the two-nucleon 
interaction that involves
a $\Delta$ isobar in the intermediate state; the existence of the
$\Delta$ isobar implies that, in the many-body system, the diagram 
Fig.~4(b) will occur (and many other diagrams involving $\Delta$s)
which---according to the above definition---represents a 
three-nucleon force (3NF).

For a spin- or isospin saturated system, the lowest order term,
Fig.~4(b), vanishes. However, when a third interaction between the
nucleon legs is introduced,
then there is a significant attractive contribution.
A very thorough calculation of diagrams of this kind has been conducted
by Dickhoff, Faessler, and M\"{u}ther.\cite{DFM82} 
Speaking in terms of the hole-line expansion, 
these authors have caculated
the three and four hole-line contributions of the ring
type with isobar degrees of freedom. 
In terms of the `nucleons only' language, the ring
diagrams involving isobars are contributions 
from many-nucleon forces.  The results
obtained by the T\"{u}bingen group\cite{DFM82} are shown in Fig.~5.

\begin{figure}[t]
\hspace*{3cm}
\epsfig{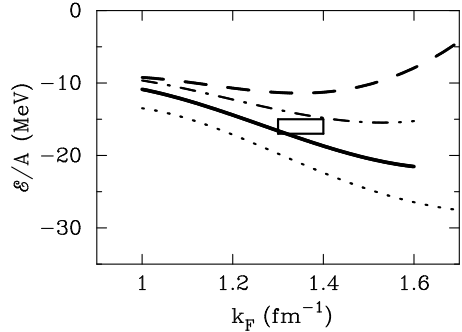}
\caption{Many-nucleon force contributions to the nuclear matter
energy. The dashed line is identical to the dashed line of
Fig.~3. Adding the 3rd and 4th order ring diagrams with 
$\Delta$-isobars results in the dash-dot curve.
The latter contributions are from the T\"ubingen group$^{37}$
who produced results for $k_F\leq 1.6$ fm$^{-1}$ only.
Further contributions from meson and nucleon renormalization
lead to the solid curve which is our final result.
The dotted curve is identical to the dotted curve of Fig.~3
and includes neither medium effects on the two-nucleon force
nor many-nucleon force contributions.}
\end{figure}

The top line (dashed) in Fig.~5, which is the starting point
of our quantitative considerations in this section, repeats the
final result
of the previous section; it is obtained in lowest order Brueckner theory
(two hole-line approximation) and includes all medium effects 
on the two-nucleon interaction discussed in Sects.~2 and 3.
Now, we add the attractive contributions from 
ring diagrams 
of third and fourth order 
involving $\Delta$ isobars 
(i.~e., 3NF and 4NF contributions)
and arrive at the dash-dot curve in Fig.~5.

Another class of diagrams that belongs into the category of 
many-nucleon forces (according to the above definition)
was calculated in Ref.~\cite{MH85} where it was shown that the selfenergy
diagrams of $\pi$ and $\rho$ together with nucleon selfenergy corrections
result in a net attractive effect (which in terms of magnitude is of the
same size as the repulsive mesonic effect discussed in Sect.~2).
When we add this contribution to the dash-dot curve in Fig.~5, then we 
obtain the solid curve which is our final result.
In this figure, we also show again the dotted curve of Fig.~3 which
represents the result when no medium effects and no many-nucleon forces
were included.
The comparison of the dotted curve with the solid curve in Fig.~5 
reveals that the sum of all medium effects on the $NN$ force
plus all many-nucleon force contributions
is much smaller than the individual contributions;
or in other words, there are large cancelations.
However, the cancelations are not perfect
and the net effect is moderately repulsive.

To summarize this and the previous section,
isobar degrees of freedom have essentially {\it two\/} consequences in
nuclear matter:
\begin{itemize}
\item {\it medium effects\/} on the two-nucleon interaction and
\item {\it many-nucleon force\/} contributions.
\end{itemize}
Both are {\it large\/} effects/contributions ---
 but, of {\it opposite\/} sign. In a consistent treatment
of degree(s) of freedom either both effects occur simultaneously,
or none.
One of these two effects alone, in isolation, does not
exist in reality.
 Therefore, to take into account only one of them
(for instance, only the three-body force contributions,
ignoring the medium effects on the corresponding two-body
diagrams)
yields a substantially distorted picture.
In fact, the almost cancelation between these two effects/contributions
may be the deeper reason why, ultimately, many-body
forces may not play a great role in nuclear physics;
it may also be the reason why the traditional two-body
force picture is by and large rather successful.

We note that our findigs in nuclear matter are supported by accurate and
independent calculations conducted for the triton binding energy.
Triton calculations of this kind were first performed
by the Hannover group.\cite{HSS83} Recently, these
calculations have been improved and extended by Picklesimer
and coworkers\cite{Pic92} using the Argonne $V_{28}$ 
$\Delta$-model.\cite{WSA84}
They find an attractive contribution to the triton energy
of --0.66 MeV from the 3NF diagrams created by the $\Delta$,
and a repulsive contribution of +1.08 MeV from the dispersive effect
on the two-nucleon force involving $\Delta$ isobars.
The total result is {\it 0.42 MeV repulsion}.\cite{Pic92}
This means that---also in the triton---the final result 
is a moderate repulsion,
similar to our findings in nuclear matter.
The consistency of the results for the two very different
many-nucleon systems suggests that there may be some general validity and
model-independence to our findings.

Besides the $\Delta$-isobar, there are other mechnisms that can
give rise to three-nucleon forces. In recent years, it has become
fashionable to consider chiral Lagrangians for nuclear
interactions. Such Langrangians may create diagrams which
represent effective three-nucleon potentials. 
However, Weinberg has shown that all these diagrams cancel.\cite{Wei90}
Moreover, based upon power counting arguments, one can show that
chiral $n$-nucleon forces with $n\geq 4$ are negligible.\cite{Kol94}

\section{Comparison to Simple Models}

\begin{figure}[t]
\hspace*{3cm}
\epsfig{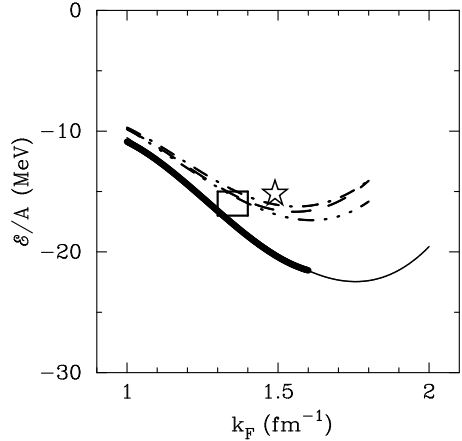}
\caption{Nuclear matter predictions by some recent two-nucleon potentials.
The dashed, dash-dot, and dash-triple-dot curves are produced by the (local)
Argonne $V_{18}$,$^{44}$ Nijm-II,$^{43}$ and Reid93$^{43}$ 
potentials, respectively; and the thin solid
line is predicted by the (nonlocal) CD-Bonn$^{45}$ potential.
The thick solid curve is identical to the solid curve
of Fig.~5 and is shown for comparison.
Note that this curve is given only for $k_F\leq 1.6$ fm$^{-1}$,
where it cannot be distinguished from the CD-Bonn curve on the scale
of the figure.
The star is the saturation point predicted by Brueckner and
Gammel$^{9}$ in 1958.
All calculations are based upon the two hole-line approximation
using the continuous choice.}
\end{figure}

In Sections 2 to 4, we have presented the results of the most
comprehensive and consistent  nuclear matter Brueckner calculations
ever conducted in terms of hadronic degrees of freedom---with
the final result given by the solid curve in Fig.~5.
After all this work which took the research groups
involved about a decade, one may raise a very naive, but reasonable
question: how does this final result compare with the predictions
from the usual simplistic approach in which a static (energy-independent)
two-nucleon potential is applied?

To answer this question, we show in Fig.~6 the saturation curves generated
by some $NN$ potentials together with the `final result' of Sect.~4
(thick solid curve, identical to solid curve of Fig.~5; note also
that Figs.~5 and 6 have slightly different scales).
The $NN$ potentials applied in Fig.~6 are all from the new 
family of high-precision/high-quality potentials developed in the
1990's.\cite{Sto94,WSS95,MSS96}
These new potentials fit the $NN$ data below 350 MeV laboratory
energy with the `perfect' $\chi^2$/datum $\approx 1$;
consequently, all these new potentials produce essentially
identical predictions for the two-nucleon observables.
However, this does not imply that these potentials are identical.
Indeed, they differ in their off-shell behavior as indicated by the
fact that some of these potentials are local and some nonlocal.

The Argonne $V_{18}$,\cite{WSS95} Nijm-II,\cite{Sto94} 
and Reid93\cite{Sto94} potentials are local and
their nuclear matter predictions are represented in Fig.~6 
by the dashed, dash-dot, and dash-triple-dot curves, respectively.
The CD-Bonn potential\cite{MSS96} is nonlocal and predicts the thin solid
curve.

Figure~6 clearly reveals that all local potentials lead to almost
identical nuclear matter results.
This statement is even true if one includes the result by
Brueckner and Gammel of 1958 with the local
Gammel-Thaler potential (see star in Fig.~6).
So, in terms of local potentials, there has not been much progress
in the past 40 years.

The story is different with the nonlocal CD-Bonn potential which
predicts a distinctly different result.
Moreover, the nonlocal CD-Bonn reproduces amazingly
well\cite{foot2} the final result from the thorough and comprehensive
calculations of Sects.~2 to 4.

Thus, the bottom line is that the result of the `faithful' calculation
that takes meson, nucleon and $\Delta$ degrees of freedom
consistently into account can, indeed, be reproduced by 
a two-nucleon potential. However, not by any potential, it has to be
a nonlocal one. Or in other words,
the nonlocality, which is one important element of the comprehensive
theory of Sects.~2 to 4, cannot be ignored.

Since the CD-Bonn is based upon relativistic meson exchange, it contains
the same nonlocalities as the fieldtheoretic models applied
in Sects.~2 to 4, except for the energy-dependence.
This energy-dependence of the $NN$ interaction can obviously be
neglected---the reason being the large cancelations between the medium
effects on the two-nucleon interaction 
(which are caused by the energy-dependence)
and the $n$-nucleon force contributions.
However, the other forms of nonlocality cannot be neglected.
In short: you can make it simple but not too simple.

\section{Relativistic `Dirac' Effects}
We will now explore an important effect not included
in the systematic investigations conducted up to Sect.~4.
Notice that, in the fieldtheoretic approach sketched in Sect.~2,
one can go one step further and use the Dirac
equation for single-particle motion in nuclear matter
\begin{equation}
(\not\!p-M-U)\tilde{u}({\bf p},s)=0
\end{equation}
or in Hamiltonian form
\begin{equation}
(\mbox{\boldmath $\alpha \cdot p$} + \beta M + \beta U)
\tilde{u}({\bf p},s) = \epsilon_{p} 
\tilde{u}({\bf p},s)
\end{equation}
with
\begin{equation}
U=U_{S} +\gamma^{0}U_{V}
\end{equation}
where $U_{S}$ is an attractive scalar
and $U_{V}$ the time component of a repulsive vector field.
(Notation as in Ref.~\cite{BD64}: $\beta=\gamma^{0}$,
  $\alpha^{l}=\gamma^{0}\gamma^{l}$, etc.)
The fields, $U_{S}$ and $U_{V}$,
are in the order of several hundred MeV and strongly density dependent.
The solution of Eq.\ (15) is
\begin{equation}
\tilde{u}({\bf p},s)=\sqrt{\frac{\tilde{E}_{p}+\tilde{M}}{2\tilde{M}}}
\left( \begin{array}{c}
          1\\
  \frac{\mbox{\boldmath $\sigma \cdot p$}}{\tilde{E}_{p}+\tilde{M}}
       \end{array} \right)
\chi_{s}
\end{equation}
with
\begin{equation}
\tilde{M}=M+U_{S} \, , \; \; \; \; \;
\tilde{E}_{p}=\sqrt{\tilde{M}^{2}+{\bf p}^{2}} \, ,
\end{equation}
and $\chi_{s}$ a Pauli spinor.

The Brueckner equation now reads
\begin{equation}
\tilde{G}(\tilde{z})
=\tilde{V}+
\tilde{V}\frac{Q}{\tilde{z}-h_{0}}\tilde{G}(\tilde{z})
\, .
\end{equation}
The essential difference to standard Brueckner theory is the use
of the potential $\tilde{V}$. 
As indicated by the tilde,
this meson-theoretic potential is evaluated
by using the spinors Eq.\ (18) instead of the free 
Dirac spinors applied in
scattering (and conventional Brueckner theory). Since $U_{S}$
 (and $\tilde{M}$) are strongly density dependent, 
so is the potential $\tilde{V}$.
$\tilde{M}$ decreases with density.
 The essential effect
in nuclear matter is a suppression of the
(attractive) $\sigma$-exchange; this suppression
 increases with density, providing
additional saturation. It turns out (see Fig.~7 below) that this
effect is so strongly density-dependent that
the empirical nuclear matter saturation energy {\it and\/}
density can be reproduced simultaneously.

The lowest-order energy of nuclear matter is
\begin{equation}
\frac{{\cal E}}{A}=\frac{1}{A}
\sum_{m\leq k_{F}}
\frac{\tilde{M}}{\tilde{E_{m}}}
\langle m | \mbox{\boldmath $\gamma \cdot p_{m}$}
+M|m\rangle 
+ \frac{1}{2A} \sum_{m,n\leq k_{F}}
\frac{\tilde{M}^{2}}{\tilde{E}_{m} \tilde{E}_{n}}
\langle mn|\tilde{G}(\tilde{z})|mn-nm\rangle -M \, .
\end{equation}
The single-particle potential,
\begin{equation}
U(m)=\frac{\tilde{M}}{\tilde{E_{m}}}
\langle m| U | m \rangle =\frac{\tilde{M}}{\tilde{E_{m}}}
 \langle m|U_{S}+\gamma^{0}U_{V}|m\rangle
=\frac{\tilde{M}}{\tilde{E_{m}}}
 U_{S} +U_{V} \, ,
\end{equation}
is determined from the $\tilde{G}$-matrix in formally the usual way,
\begin{equation}
U(m)=Re\sum_{n\leq k_{F}}\frac{\tilde{M}^{2}}{\tilde{E_{n}}\tilde{E_{m}}}
 \langle mn|\tilde{G}(\tilde{z})|mn-nm\rangle
\end{equation}
with $m$ below and above the Fermi surface (continuous choice)
and
\begin{equation}
\tilde{z}=\epsilon_m+\epsilon_n \, ,
\end{equation}
where the single particle energy is given by
\begin{eqnarray}
\epsilon_{m}&=&\frac{\tilde{M}}{\tilde{E_{m}}}
\langle m| \mbox{\boldmath $\gamma \cdot p_{m}$} + M |m\rangle
+ U(m)\\
 & = & \tilde{E_{m}} + U_{V} \, .
\end{eqnarray}

Note that, in this approach, the nucleon states 
$|m\rangle$ and $|n\rangle$ are
represented by Dirac spinors of the kind Eq.\ (18)
and an appropriate isospin wavefunction,
$\langle m|$ and $\langle n|$ are the adjoint Dirac spinors
$\bar{\tilde{u}}=\tilde{u}^{\dagger}\gamma^{0}$;
$\tilde{E}_{m}=\sqrt{\tilde{M}^{2}+{\bf p}^{2}_{m}}$.
The normalization of the Dirac spinors is $\bar{\tilde{u}} {\tilde u} = 1$.

Results are shown in Fig.~7 where the lower solid and lower dashed curves
repeat the predictions presented in the previous section
for the Argonne $V_{18}$ and the CD-Bonn potentials, respectively.
To these curves, we now add the relativistic Dirac effects
based upon the formalism just sketched and calculated in 
Ref.~\cite{BM90}. This results in the upper dashed and solid curves for
$V_{18}$ and CD-Bonn, respectively.

As discussed, this Dirac effect allows a correct reproduction
of the saturation density and energy---as demonstrated by the upper
CD-Bonn curve.
It is also evident that slight overbinding in the nonrelativistic
approach is a necessary
prerequisite for ultimately predicting density and energy right.
As revealed by the upper dashed line in Fig.~7, local potentials
lack binding energy and, therefore, cannot reproduce nuclear
saturation correctly.

Alternatively, one may take a phenomenological approach in which
one tries to simulate the Dirac effect by a repulsive 3NF.
This method is pursued by the Urbana group\cite{APR98}
and we have included their results in Fig.~7 in terms
of the dotted lines. The lower dotted curve is the variational result
with the $V_{18}$ two-nucleon potential and should be compared
with the lower dashed curve (Brueckner with $V_{18}$).
The closeness of these two curves confirms what we discussed in
conjunction with Fig.~1, namely, a satisfactory agreement
between Brueckner theory and the variational approach.
The Urbana group developed a repulsive phenomenological 3NF
designed to improve the nuclear saturation prediction.
Adding this 3NF leads us from the lower dotted curve
to the upper dotted curve. The latter is obviously
close to the upper dashed line which implies that the repulsive 3NF
is simulating the Dirac effects well.
The upper dotted and dashed curves are the final results for
Argonne $V_{18}$ and prove that---no matter what mechanisms
are invoked to get the saturation density right---local
potentials always underbind nuclear matter.

\begin{figure}[t]
\hspace*{3cm}
\epsfig{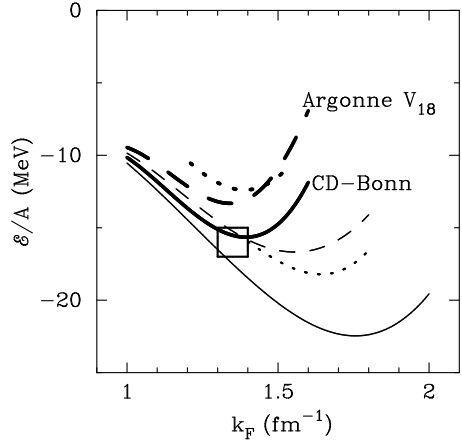}
\caption{The impact of relativistic Dirac effects on nuclear matter
saturation. The lower solid and dashed curves are without
and the corresponding upper curves are with Dirac effects.
Dashed curves represent Brueckner results based upon the 
Argonne $V_{18}$ potential and
solid curves correspond to CD-Bonn. Dotted curves are variational
results with $V_{18}$ where the upper dotted curve includes
a repulsive 3NF (from Akmal {\it et al.}$^{49}$).}
\end{figure}

In conclusion, nonlocality of the nuclear force is crucial
to explain nuclear binding quantitatively.
This should be no surprise since the
nuclear force is inherently nonlocal, anyhow.

\section{Conclusions}
Using Brueckner theory,
we have investigated systematically the influence of nonnucleonic
degrees of freedom on the predicted properties of nuclear matter.\cite{foot3}
In general, many-body calculations of this kind can be very involved
and most cumbersome. However, in the framework of Brueckner theory,
these rather sophisticated investigations can be conducted
in a straightforward and managable way.
This is one of the great advantages of Brueckner theory.

One result of our investigation is that explicit meson and
$\Delta$-isobar degrees of freedom do not improve the saturation
properties of nuclear matter. However, they teach us something
about many-body forces: a mechnism that has the potential
of creating many-nucleon forces should be treated consistently
in the two- and many-nucleon system. Even though individual
contributions caused by this mechanism can
be large, the net result turns out to be very small---due
to characteristic cancelations between effects from
two- and many-nucleon terms.
Indeed, the final result is very similar to what is obtained
from a {\it nonlocal\/} two-nucleon force, since nonlocality
is an inherent quality of the more comprehensive and
fundamental fieldtheoretic approach.
If the force is local, 
insufficient binding is predicted.

We also discussed an extension of Brueckner theory in which
the relativistic Dirac equation is used for single particle
motion (also known as Dirac-Brueckner-Hartree-Fock). 
This has a substantial impact
on nuclear saturation such that the empirical density and
energy can be reproduced.

Based upon the work summarized in this talk and the countless papers
not mentioned,  we conclude
that Brueckner theory has contributed in a crucial way to
our understanding of the nuclear many-body problem.

\section*{Acknowledgments}
This work was supported in part by the U.S.~National Science Foundation
under Grant No.~PHY-9603097.


\begin{thebibliography}{99}
\bibitem{Eul37} H. Euler, {\it Z. Physik} {\bf 105}, 553 (1937).
\bibitem{Wei35} C. F. von Weiz\"{a}cker, {\it Z. Physik} {\bf 96}, 
431 (1935).
\bibitem{Jas51} R. Jastrow, {\it Phys.\ Rev.} {\bf 81}, 165 (1951).
\bibitem{BLM54} K. A. Brueckner {\it et al.},
{\it Phys.\ Rev.\/} {\bf 95}, 217 (1954).
\bibitem{Bru54} K. A. Brueckner, {\it Phys.\ Rev.\/} {\bf 96}, 508 (1954);
{\it ibid.} {\bf 100}, 36 (1955).
\bibitem{BL55} K. A. Brueckner and C. A. Levinson,
{\it Phys.\ Rev.\/} {\bf 97}, 1344 (1955).
\bibitem{Wat53} K. M. Watson, {\it Phys.\ Rev.} {\bf 89}, 575 (1953).
\bibitem{Gol57} J. Goldstone, {\it Proc.\ Roy.\ Soc.\ (London)}
{\bf A239}, 267 (1957).
\bibitem{BG58} K. A. Brueckner and J. L. Gammel,
{\it Phys.\ Rev.\/} {\bf 109}, 1023 (1958).
\bibitem{GT57} J. L. Gammel and R. M. Thaler, {\it Phys.\ Rev.\/}
{\bf 107}, 291, 1339 (1957).
\bibitem{Day67} B. D. Day, {\it Rev.\ Mod.\ Phys.} {\bf 39},
719 (1967).
\bibitem{RB67} R. Rajaraman and H. A. Bethe, {\it Rev.\ Mod.\ Phys.}
{\bf 39}, 745 (1967).
\bibitem{Bra67} B. H. Brandow, {\it Rev.\ Mod.\ Phys.\/}
{\bf 39}, 771 (1967).
\bibitem{Bet71} H. A. Bethe, {\it Ann.\ Rev.\ Nucl.\ Sci.} {\bf 21},
93 (1971).
\bibitem{Sie70} P. J. Siemens, {\it Nucl.\ Phys.} {\bf A141}, 225 (1970).
\bibitem{HT70} M. I. Haftel and F. Tabakin, {\it Nucl.\ Phys.\/}
{\bf A158}, 1 (1970).
\bibitem{Coe70} F. Coester {\it et al.},
{\it Phys.\ Rev.\ C\/} {\bf 1}, 769 (1970).
\bibitem{Spr72} D. W. L. Sprung, {\it Adv.\ Nucl.\ Phys.}
{\bf 5}, 225 (1972).
\bibitem{Cla75} J. W. Clark, {\it Crisis in Nuclear-Matter Theory},
unpublished (1975); 
{\it Nucl.\ Phys.} {\bf A328}, 587 (1979);
{\it Prog.\ Part.\ Nucl.\ Phys.} {\bf 2}, 89 (1979).
\bibitem{Day78} B. D. Day, {\it Rev.\ Mod.\ Phys.\/} {\bf 50}, 495 (1978).
\bibitem{Day81} B. D. Day, {\it Phys.\ Rev.\ Lett.\/} {\bf 47}, 226 (1981);
{\it Phys.\ Rev.\ C\/} {\bf 24}, 1203 (1981).
\bibitem{PW79} 
V. R. Pandharipande and R. B. Wiringa,
{\it Rev.\ Mod.\ Phys.} {\bf 51}, 821 (1979).
\bibitem{FR75} 
S. Fantoni and S. Rosati, 
{\it Nuovo Cim.} {\bf 25A},
593 (1975); {\it ibid.} {\bf 43A}, 413 (1978);
O. Benhar {\it et al.},
{\it Nucl.\ Phys.} {\bf A328}, 127 (1979).
\bibitem{MB1} 
Proc. Int. Conf. on Recent Progress in Many-Body Theories, 
Trieste, 1978,
{\it Nucl.\ Phys.} {\bf A328} (1979).
\bibitem{LP81} R. B. Wiringa and V. R. Pandharipande, 
{\it Phys.\ Lett.} {\bf B99}, 1 (1981);
I. E. Lagaris and V. R. Pandharipande, 
{\it Nucl.\ Phys.} {\bf A359}, 349 (1981).
\bibitem{DW85} B. D. Day and R. B. Wiringa, {\it Phys.\ Rev.\ C\/}
{\bf 32}, 1057 (1985).
\bibitem{MB4} 
Proc.\ Fourth Int.\ Conf.\ on Recent Progress in Many-Body Theories, 
San Francisco, 1985, unpublished.
\bibitem{WSA84} R. B. Wiringa {\it et al.},
{\it Phys.\ Rev.\ C\/} {\bf 29}, 1207 (1984).
\bibitem{foot1} 
The single particle potential for a nucleon $m$ below the Fermi
surface is defined in terms of the
`on-shell' Brueckner $G$ matrix by
\begin{equation}
U(m)=Re \sum_{n\leq k_{F}}
 \langle mn|{G(w)}|mn-nm\rangle 
\end{equation}
with starting energy $w=e_m+e_n$, where $e_m$ and $e_n$ denote
the nonrelativistic single particle energies 
of nucleon $m$ and $n$, respectively.
If this same definition is also applied to nucleons above
the Fermi surface, we speak of the {\it continuous choice}.
If, however, $U\equiv 0$ is assumed above $k_F$, then we
have the {\it conventional or gap choice.} 
\bibitem{Son98} H. Q. Song, M. Baldo, G. Giansiracusa, and U. Lombardo,
{\it Phys.\ Rev.\ Lett.} {\bf 81}, 1584 (1998); M. Baldo, private communication.
\bibitem{JLM76} J. P. Jeukenne, A. Lejeune, and C. Mahaux,
{\it Phys.\ Reports\/} {\bf 25}, 83 (1976).
\bibitem{GL79} P. Grang\'{e} and A. Lejeune, {\it Nucl.\ Phys.} {\bf A327},
335 (1979).
\bibitem{Mah79} C. Mahaux, {\it Nucl.\ Phys.} {\bf A328}, 24 (1979).
\bibitem{Sch74} D. Sch\"{u}tte, {\it Nucl.\ Phys.} {\bf A221}, 450 (1974).
\bibitem{Sch61} S. S. Schweber, {\it An Introduction to Relativistic
Quantum Field Theory} (Row, Peterson and Co.; Evanston, Ill., U.S.A.; 1961)
Chapters 11 and 13.
\bibitem{MHE87} R. Machleidt, K. Holinde, and C. Elster,
{\it Phys.\ Reports\/} {\bf 149}, 1 (1987).
\bibitem{DFM82} W. H. Dickhoff, A. Faessler, and H. M\"{u}ther,
{\it Nucl.\ Phys.} {\bf A389}, 492 (1982).
\bibitem{MH85} R. Machleidt and K. Holinde, {\it Phys.\ Lett.}
{\bf 152B}, 295 (1985).
\bibitem{HSS83} C. Hajduk, P. U. Sauer, and W. Strueve,
{\it Nucl.\ Phys.} {\bf A405}, 581 (1983).
\bibitem{Pic92} A. Picklesimer, R. A. Rice, and R. Brandenburg, 
{\it Phys.\ Rev.\ C\/} {\bf 45}, 2045, 2624 (1992); 
{\it ibid.} {\bf 46}, 1178 (1992).
\bibitem{Wei90} S. Weinberg, {\it Phys.\ Lett.} {\bf B251}, 288 (1990);
{\it Nucl.\ Phys.} {\bf B363}, 3 (1991);
{\it Phys.\ Lett.} {\bf B295}, 114 (1992).
\bibitem{Kol94} U. van Kolck, {\it Phys.\ Rev.\ C\/} {\bf 49}, 2932 (1994).
\bibitem{Sto94} 
V. G. J. Stoks {\it et al.}, 
{\it Phys.\ Rev.\ C\/} {\bf 49}, 2950 (1994).
\bibitem{WSS95} R. B. Wiringa, V. G. J. Stoks, and R. Schiavilla,
{\it Phys.\ Rev.\ C\/} {\bf 51}, 38 (1995).
\bibitem{MSS96} R. Machleidt, F. Sammarruca, and Y. Song,
{\it Phys.\ Rev.\ C\/} {\bf 53}, 1483 (1996).
\bibitem{foot2} We note that the very close agreement between the CD-Bonn
curve (thin solid line) and the thick solid curve in Fig.~6 is,
of course, accidental. However, the agreement within $\pm 1$ MeV
is not.
\bibitem{BD64} J. D. Bjorken and S. D. Drell, {\it Relativistic
Quantum Mechanics}, McGraw-Hill, New York (1964).
\bibitem{BM90} R. Brockmann and R. Machleidt, {\it Phys.\ Rev.\ C\/} 
{\bf 42}, 1965 (1990).
\bibitem{APR98} A. Akmal {\it et al.},
{\it Phys.\ Rev.\ C\/} {\bf 58}, 1804 (1998).
\bibitem{foot3} A comprehensive and more detailed review of the topics
discussed in this talk and the pertinent literature
can be found in sections 9 and 10 
of Ref.~\cite{Mac89} which is also the original reference for many
of the results presented.
\bibitem{Mac89} R. Machleidt, {\it Adv.\ Nucl.\ Phys.} {\bf 19}, 189 (1989).
\end{thebibliography}
\end{document}